\documentclass[aps,prd,reprint,superscriptaddress,nofootinbib]{revtex4-1}
\usepackage{graphicx}
\usepackage{dcolumn}
\usepackage{bm}
\usepackage{epstopdf}
\usepackage{amsmath}
\usepackage{hyperref}
\usepackage{color}
\usepackage{cancel}

\begin{document}
\raisebox{20pt}[0pt][0pt]{\hspace*{35mm}  RIKEN-iTHEMS-Report-22}

\title{Attractive $N$-$\phi$ interaction and two-pion tail from lattice QCD near physical point }

\author{Yan Lyu
}
\email{helvetia@pku.edu.cn}
\affiliation{State Key Laboratory of Nuclear Physics and Technology, School of Physics, Peking University, Beijing 100871, China }
\affiliation{Interdisciplinary Theoretical and Mathematical Sciences Program (iTHEMS), RIKEN, Wako 351-0198, Japan}
\author{Takumi Doi
}
\email{doi@ribf.riken.jp}
\affiliation{Interdisciplinary Theoretical and Mathematical Sciences Program (iTHEMS), RIKEN, Wako 351-0198, Japan}
\author{Tetsuo Hatsuda
}
\email{thatsuda@riken.jp}
\affiliation{Interdisciplinary Theoretical and Mathematical Sciences Program (iTHEMS), RIKEN, Wako 351-0198, Japan}
\author{Yoichi Ikeda
}
\email{yikeda@cider.osaka-u.ac.jp}
\affiliation{Center for Infectious Disease Education and Research, Osaka University, Suita 565-0871, Japan}
\author{Jie Meng
}
\email{mengj@pku.edu.cn}
\affiliation{State Key Laboratory of Nuclear Physics and Technology, School of Physics, Peking University, Beijing 100871, China }
\affiliation{Yukawa Institute for Theoretical Physics, Kyoto University, Kyoto 606-8502, Japan}
\author{Kenji Sasaki
}
\email{kenjis@cider.osaka-u.ac.jp}
\affiliation{Center for Infectious Disease Education and Research, Osaka University, Suita 565-0871, Japan}
\author{Takuya Sugiura
}
\email{takuya.sugiura@riken.jp}
\affiliation{Interdisciplinary Theoretical and Mathematical Sciences Program (iTHEMS), RIKEN, Wako 351-0198, Japan}

\date{\today}
\begin{abstract}
First results on the interaction between the $\phi$-meson and the nucleon ($N$)  are presented based on the ($2+1$)-flavor lattice QCD simulations with nearly physical quark masses.
Using the HAL QCD method, the spacetime correlation of the $N$-$\phi$ system in the spin 3/2 channel 
is converted into the  $N$-$\phi$ scattering phase shift through the interaction potential.
The $N$-$\phi$   potential appears to be a combination of a  short-range attractive core 
and a long-range attractive tail. The latter is found to be consistent with the two-pion exchange (TPE)
 obtained from the interaction between a color-dipole and the nucleon.
The resultant scattering length and effective range for $m_{\pi}=$ 146.4 MeV are $ a^{(3/2)}_0=-1.43(23)_{\rm stat.}\left(^{+36}_{-06}\right)_{\rm syst.} {\rm fm}$ and  $ r^{(3/2)}_{\rm eff}=2.36(10)_{\rm stat.}\left(^{+02}_{-48}\right)_{\rm syst.} {\rm fm}$, respectively.
The magnitude of the scattering length is shown to have nontrivial dependence of $m_{\pi}$ and is sensitive to the existence of the long-range tail from TPE.
\end{abstract}


\maketitle



\section{Introduction}
The interaction between vector mesons and the nucleon ($N$) is one of the most fundamental quantities to study the meson properties in nuclear matter 
~\cite{Hayano-Hatsuda2010}.
Among others, the $\phi$-meson attracts particular interest, since the interactions of the $s\bar{s}$ pair with the nucleon and nucleus are not well understood at low energies. 
 
Experimentally, the production and absorption of the $\phi$-meson  in nuclear matter has been actively studied through photon and proton($p$)-induced reactions with various nuclear targets
(Spring8-LEPS~\cite{Ishikawa2005}, KEK-PS E325~\cite{KEK-PS2007}, JLab-CLAS~\cite{CLAS2010}, and COSY-ANKE~\cite{Polyanskiy2011}) as summarized in Ref.~\cite{Tolos2020}.
Recently, the ALICE Collaboration at LHC presented a new result on the $p$-$\phi$ interaction  through the femtoscopic analysis of the $p$-$\phi$ pair produced in the $pp$ reaction:
They reported a spin-averaged scattering length $a_{p\phi} = -0.85(34)_{\rm stat.}(14)_{\rm syst.}$ fm~\cite{ALICE2021}, an order of magnitude larger than that obtained from the photoproduction data by the CLAS Collaboration at JLab~\cite{Dey:2014tfa} combined with the vector meson dominance \cite{Strakovsky2020}.
            
Theoretically, the properties of the $\phi$-meson in nuclear matter and its relation to the in-medium $\bar{s}s$ condensate have been extensively discussed~\cite{Hatsuda:1991ez,Asakawa:1994tp,Gubler:2016itj,Gubler:2018ctz}.  
The microscopic origin of the force between $\phi$ and $N$ is also an important open problem.
In particular, the two-pion exchange (TPE) as the major long-range contribution between a color-dipole and the nucleon~\cite{TarrusCastella:2018php} is of strong relevance to this problem and will shed new light on the important role of pion dynamics in hadron interactions.

 Under these experimental and theoretical circumstances, it is most desirable to carry out realistic lattice QCD simulations of the $N$-$\phi$ interaction.
In this paper, we report a first simulation of the $N$-$\phi$ system in a large lattice volume $\simeq$ (8.1 fm)$^3$  with light dynamical quarks near the physical point.
We focus on the highest-spin $N$-$\phi$ system, the $^4S_{3/2}$ channel,  with the notation $^{2s+1}L_J$ ($s$ = total spin, $L$ = orbital angular momentum, $J$ = total angular momentum). This is because  its coupling to two-body open channels $\Lambda K(^2D_{3/2})$ and $\Sigma K(^2D_{3/2})$ are kinematically suppressed at low energies  due to their $D$-wave nature, similar to the situation for the highest-spin   $N$-$\Omega$ system (the $^5S_{2}$ channel)  \cite{Iritani2019PLB}. 
Moreover, decay processes with three or more particles in the final state such as $\Sigma \pi K$, $\Lambda \pi K$, and $\Lambda \pi\pi K$ are expected to be suppressed due to the small phase space.
Indeed, we will show below that such decays are not visible from our lattice data in the $^4S_{3/2}$  channel.
By using the HAL QCD method~\cite{Ishii2007, Ishii2012, Aoki2020}, which converts the spacetime correlation of hadrons to the physical observables,  
we obtain the $N$-$\phi$ scattering phase shift, the $S$-wave scattering length $a_{0}^{(3/2)}$ and the effective range $r_{\rm eff}^{(3/2)}$ in the $^4S_{3/2}$ channel. (For recent applications of the HAL QCD method to the baryon-baryon interactions, see Refs.~\cite{Gongyo2018, Iritani2019PLB, Sasaki2020, Lyu2021}.)

This paper is organized as follows. Section~\ref{sec-II} provides a brief review of the HAL QCD method. Details of our lattice setup are given in Sec.~\ref{sec-III}.  Numerical results on the $N$-$\phi$ potential and scattering properties are presented in Sec.~\ref{sec-IV}. 
Section~\ref{sec-V} is devoted to the summary.  
The detailed systematic analysis on the $N$-$\phi$ potential is given in Appendix~\ref{sec-VI} and~\ref{sec-VII}. 
 
\section{HAL QCD Method}\label{sec-II}
Let us consider normalized correlation function of interacting $N$ and $\phi$ 
 as a function of the spatial coordinate $\bm{r}$ and the Euclidean time $t$ ~\cite{Ishii2007}, 
 \begin{eqnarray}\label{eq-R}
    R(\bm{r},t)&=&
   \frac{\sum_{\bm{x}}\langle 0| N(\bm{r}+\bm{x},t)\phi(\bm{x},t)\overline{\mathcal{J}}(0)|0\rangle} 
    {\sqrt{Z_{\phi}Z_N}  e^{-(m_{N}+m_{\phi})t}} \nonumber \\
    &=&\sum_n a_n \psi_n(\bm{r})e^{-(\Delta E_n)t}+O(e^{-(\Delta E^*)t}).
\end{eqnarray}
Here the energy eigenvalue and the energy shift for  the  elastic scattering state of $N$-$\phi$  are given by 
$E_n=\sqrt{m_N^2+\bm{k}_n^2}+\sqrt{m_\phi^2+\bm{k}_n^2}$ 
and   $\Delta E_n=E_n-(m_N+m_\phi)$, respectively, with $\bm{k}_n$ being the relative momentum  in the center of mass frame.  
The equal-time Nambu-Bethe-Salpeter (NBS) wave function is denoted by $\psi_n(\bm r)$.  
The wave function renormalization constant for $N$ ($\phi$) is  given by $Z_{N}$ ($Z_{\phi}$).
 The contributions from the inelastic scattering states are exponentially suppressed as $O(e^{-(\Delta E^*)t})$ with $\Delta E^*$ being the energy of the inelastic threshold relative to $m_N+m_{\phi}$. 
The overlapping factor between the $N$-$\phi$   source operator  $\overline{\mathcal{J}}(0)$
 and the $n$th eigenstate $|n\rangle$ is given by  $a_n=\langle{n}|\overline{\mathcal{J}}(0)|0\rangle$.
We employ the wall-type source operator  with the Coulomb gauge fixing to enhance the overlap of the source with the $N$-$\phi$ scattering state.
 For the sink operators,  the following local composite operators are adopted  to define the potential,
 \begin{eqnarray}\label{eq-op}
 N_\alpha(x) &=& \epsilon_{abc}\left[{{u^a}}(x)C\gamma_5d^b(x)\right]u^c_\alpha(x),  \nonumber \\ 
  \phi_i(x) &=& \delta_{ab}\bar{s}^a(x)\gamma_is^b(x).
 \end{eqnarray}
 Here $a$, $b$, and $c$ are the color indices, $i$ is the vector index, $\alpha$ is the Dirac index restricted to the upper two components, and 
 $C=\gamma_4\gamma_2$ is the charge conjugation.
Thanks to Haag-Nishijima-Zimmermann's theorem ~\cite{Zimmermann1987}, 
 it is sufficient  to consider these local  operators to extract the 
  observables from the correlation function.
The Misner's method for approximate partial wave decomposition on a cubic grid
is used for the $S$-wave projection on the lattice~\cite{Misner:1999ab,Miyamoto2020}.
The normalized four-point functions $R(\bm{r},t)$ with the spin projection to $J=3/2$ are used to extract the potential.

As shown in  Ref.\cite{Ishii2012}, $R(\bm{r},t)$ satisfies the integrodifferential equation,
\begin{equation}\label{eq-TDHAL}
 \begin{split}
  &\left[ \frac{1+3\delta^2}{8\mu}\frac{\partial^2}{\partial t^2}-\frac{\partial}{\partial t}-H_0 +O(\delta^2\partial_t^3) \right] R(\bm{r},t) \\
 &= \int d\bm{r}'U(\bm{r},\bm{r}')R(\bm{r}',t),
 \end{split}
\end{equation}
with $H_0=-\nabla^2/(2\mu)$, $\mu=m_N m_\phi/(m_N+m_\phi)$ being the reduced mass,  $\delta=(m_N-m_\phi)/(m_N+m_\phi)$ being the
 mass asymmetry.  We neglect the term of $O(\delta^2\partial_t^3)$, since it is found to be consistent with zero within the statistical error in our simulation.  In practical calculations, a derivative expansion of the nonlocal potential is employed~\cite{OKubo-Marshak1958,Iritani2019PRD}, 
\begin{equation}
U(\bm{r},\bm{r}')=V(r)\delta(\bm{r}-\bm{r}')+ \sum\limits_{n=1}V_{n}(\bm{r})\nabla^{n}\delta(\bm{r}-\bm{r}').
\end{equation}
 Then, we obtain the the leading-order  (LO) central potential as
\begin{equation}\label{eq-V}
 V(r)= R^{-1}(\bm{r},t)\left[ \frac{1+3\delta^2}{8\mu}\frac{\partial^2}{\partial t^2}-\frac{\partial}{\partial t}-H_0 \right] R(\bm{r},t).
\end{equation}
The truncation error of the derivative expansion is found to be small at low energies by using the finite-volume spectral analysis~\cite{Iritani2019Jhep, Lyu2022} (See Appendix~\ref{sec-VI}). 
We note that the NBS wave function $\psi_n (\bm{r})$, whose asymptotic form 
 provides the two-body $S$-matrix, obeys the Klein-Gordon equation~\cite{Aoki2010}, 
\begin{equation}
  (\bm{k}_n^2 + \nabla^2) \psi_n(\bm{r}) =  2\mu \int d\bm{r}' U(\bm{r}, \bm{r}')   \psi_n(\bm{r}').
\end{equation}
 Therefore, we calculate the phase shift by solving this equation in the infinite volume with the LO potential in Eq.~(\ref{eq-V}).

\section{Lattice setup}\label{sec-III}
The ($2+1$)-flavor gauge configurations are generated with the Iwasaki gauge action at $\beta=1.82$ and the nonperturbatively $O(a)$-improved Wilson quark action with stout smearing at nearly physical quark masses~\cite{Ishikawa2016}.
The lattice spacing is $a\simeq0.0846$ fm ($a^{-1}\simeq2333$ MeV), and the lattice volume is $L^4=96^4$, corresponding to $La\simeq8.1$ fm.
Listed in Table ~\ref{tab-mass} together with 
 the experimental values~\cite{PDG2020} are the isospin-averaged masses of $\pi$, $K$, $\phi$, and $N$ obtained from 
  the single-state fitting  in the interval $t/a=15$$-$25 (mesons)  and 11$-$18 (the nucleon) 
  with the statistical errors in the parentheses.
The lattice results are about 6\%  larger than the experimental values for $\pi$ and $K$, and about 3\% (2\%) larger for $\phi$ ($N$).
Since our quark masses are slightly heavier than the physical masses, the decay $\phi \rightarrow K\bar{K}$ is kinematically forbidden, while
the $\phi \rightarrow 3 \pi$ is  allowed.  We found that the effective mass $m_{\phi}(t) $ is stable  even up to $t/a$ =40, so that
  $\phi$ behaves as an approximate asymptotic state.
\begin{table}[htbp]
\caption{ Isospin-averaged hadron masses with statistical errors obtained from (2+1)-flavor lattice QCD simulations together with the experimental values.
}
\begin{tabular}{ccc}
  \hline\hline
    Hadron &~~Lattice [MeV]  &~~Expt. [MeV] \\
  \hline
    $\pi$ &\ \ 146.4(4)  & \ \ 138.0\\
    $K$  & \ \ 524.7(2) & \ \ 495.6\\
    $\phi$ &\ \ 1048.0(4) &\ \ 1019.5\\
    $N$  &\ \ 954.0(2.9)  & \ \ 938.9 \\
  \hline\hline
\end{tabular} \label{tab-mass}
\end{table}

We use 200 gauge configurations separated by 10 trajectories.
 To reduce the statistical fluctuation, the forward and backward propagations are averaged in each configuration,
  the hypercubic symmetry on the lattice (four rotations) is utilized, and 80 measurements are performed by shifting the source position in a temporal direction.
In total,  128,000 measurements were taken.
 Quark propagators are calculated by the domain-decomposed solver~\cite{Ishikawa2021} with the  periodic boundary condition for all directions.
 Hadron correlation functions are obtained by the unified contraction algorithm~\cite{Doi2013}.
The OZI (Okubo-Zweig-Iizuka) violating $s\bar{s}$ annihilation  is not considered. 
The statistical errors are evaluated by the jackknife method with a bin size of 20 configurations throughout this paper, and a comparison with a bin size of 40 configurations shows that the bin size dependence is small.
The major systematic error stems from the variation of the potential with respect to $t/a$ as discussed below.

\begin{figure}[htbp]
  \centering
  \includegraphics[width=8cm]{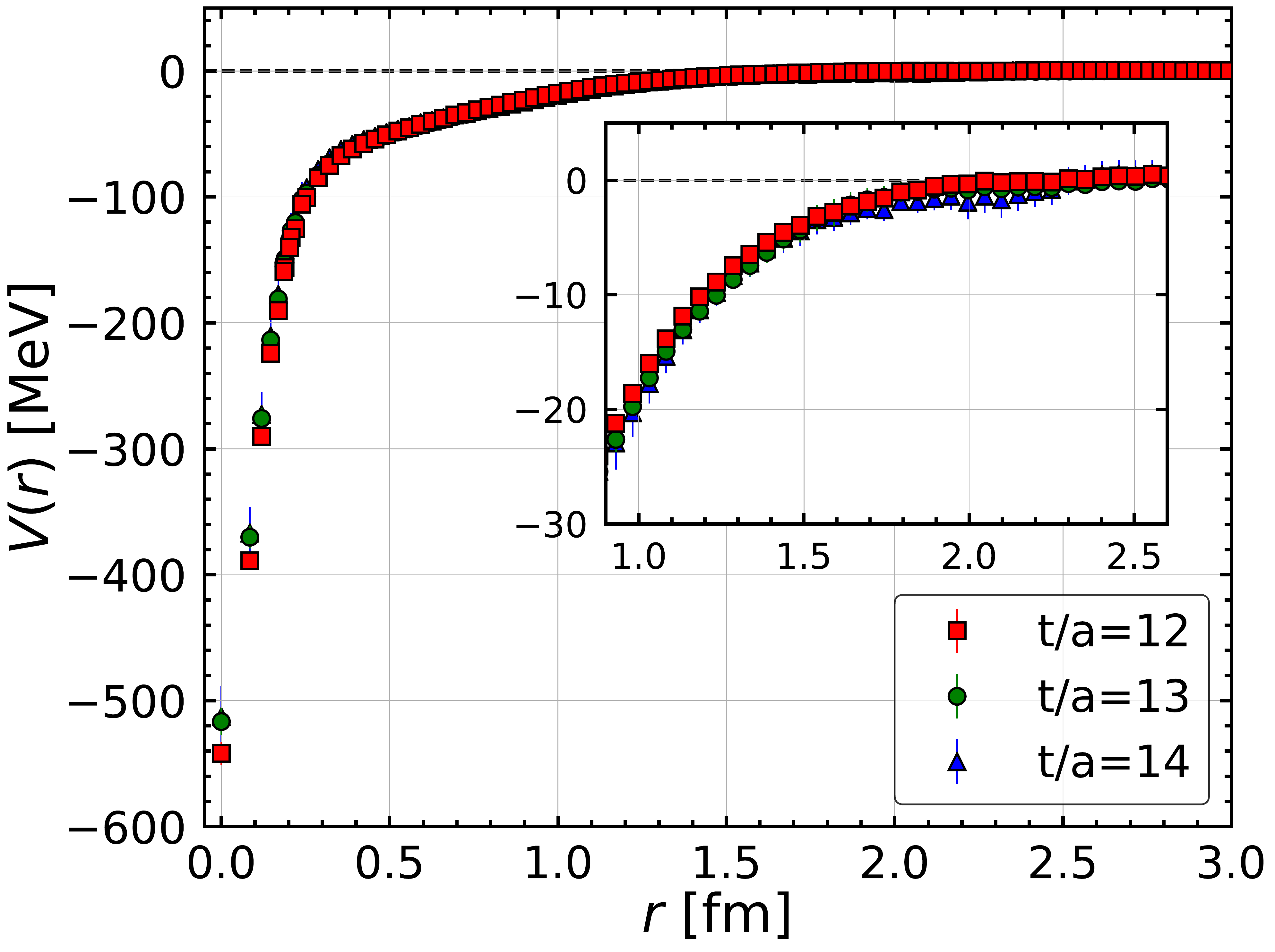}
  \caption{(Color online).
  The $N$-$\phi$ potential $V(r)$ in the ${^4S_{3/2}}$ channel as a function of separation $r$ at
 Euclidean time $t/a= 12$ (red squares), $13$ (green circles), and $14$ (blue triangles).
  } \label{Fig1}
\end{figure}

\section{Numerical results}\label{sec-IV}
The $N$-$\phi$ potential $V(r)$ in the $^4S_{3/2}$ channel defined in Eq.~(\ref{eq-V}) with the
lattice measurement of $R(\bm r, t)$ is shown in Fig.~\ref{Fig1} for Euclidean times, $t/a=12$, $13$, and $14$. (See Appendix~\ref{sec-VII} for the $t$ dependence of $V(r)$ in a wider range of $t$.)
These Euclidean times are chosen such that they are large enough to suppress contaminations from excited states in the single-hadron correlator and simultaneously small enough to avoid exponentially increasing statistical errors.
The variation of the potential between different $t/a$ is due to the contamination of inelastic states and the truncation of the derivative expansion.
Such a variation is taken into account as a major source of the systematic error in our final results.
 A relatively small variation of $V(r)$  as a function of $t/a$  indicates 
  that the  $N$-$\phi$ correlation function is mostly dominated by the elastic scattering states in the $^4S_{3/2}$ channel without 
  significant effects from
   the two-body open channels  ($\Lambda K(^2D_{3/2})$ and $\Sigma K(^2D_{3/2})$) and the  three-body open channels including $N\phi \rightarrow \{\Sigma^* K, \Lambda(1405) K\} \rightarrow \{\Lambda \pi K, \Sigma \pi K$\}.
This is in sharp contrast to the  $^2S_{1/2}$ case where we found that the $N$-$\phi$ potential shows a clear $t$ dependence, as expected from the $S$-wave fall-apart decay into $\Lambda K(^2S_{1/2})$ and $\Sigma K(^2S_{1/2})$.

The potential $V(r)$ in the $^4S_{3/2}$ channel shown in Fig.~\ref{Fig1} is attractive for all distances and has a characteristic two-component structure, the 
attractive core at short distance and the attractive tail at long distance,  similar to the case of the $N\Omega(^5S_{2})$ potential~\cite{Iritani2019PLB}.  We note that the Pauli exclusion principle between quarks, which partially gives rise to the repulsive core in the $NN$ interaction~\cite{Inoue2012,Oka2000},  does not operate in the present case, since $N$ and $\phi$ have no common valence quarks.  
 
\begin{figure}[htbp]
  \centering
  \includegraphics[width=8cm]{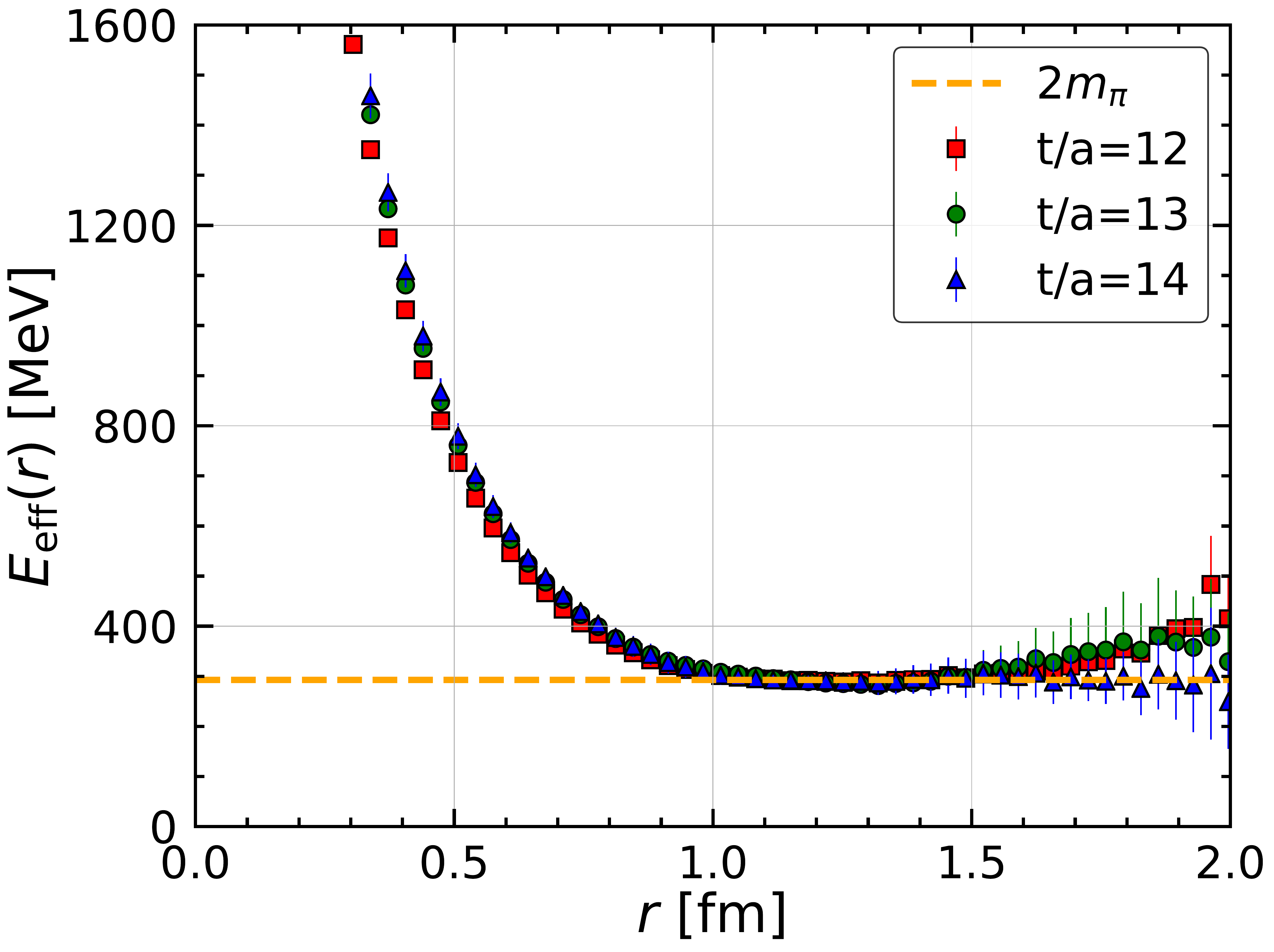}
  \caption{(Color online).
  The spatial effective energy $E_{\rm eff}(r)$ as a function of separation $r$ at
 Euclidean time $t/a= 12$ (red squares), $13$ (green circles) and $14$ (blue triangles).
The orange dashed line corresponds to $2m_\pi$ with lattice pion mass $m_\pi=146.4$ MeV.
  } \label{Fig2}
\end{figure}

As has been discussed  for the interaction between color dipoles~\cite{Bhanot:1979vb,Fujii:1999xn,Brambilla:2015rqa},
  nonperturbative gluon exchange is  expected to appear in the form of  the TPE at long distance.
  The idea was generalized to the interaction between a color-dipole and the nucleon with the result, 
 $V(r \gg (2m_{\pi})^{-1} ) = -\alpha  \frac{\exp (-2m_\pi r)}{r^{2}}$, where $\alpha$ is proportional to $m_{\pi}^4$ \cite{TarrusCastella:2018php}.  
To check such a long distance behavior of $V(r)$,  we show in Fig.~\ref{Fig2} the spatial effective energy as a function of $r$,
\begin{equation}
 E_{\rm eff}(r)=-\frac{\ln[-V(r)r^{2}/\alpha]}{r},
\end{equation}  
with $\alpha \simeq 91~{\rm MeV}\cdot{\rm fm}^{2}$ determined by fitting the lattice data of $V(r)$ at long distance.  
We find that $E_{\rm eff}(r)$ has a plateau at  $2m_\pi = 292.8$ MeV  for $r > 1.0$ fm, which indicates that the long-range part of the $N$-$\phi$ potential is indeed dominated by the TPE.

In order to convert the potential to physical observables, we perform an uncorrelated fit of the lattice QCD potential 
by using two different functional forms,
\begin{eqnarray}\label{eq-fit}
 \! \! {\rm{A}}: V_{\rm{fit}}(r) &=& \sum_{i=1,2}a_i e^{-(r/b_i)^2} + a_3 m_\pi^4 f(r;b_3)  \frac{e^{-2m_\pi r}}{r^{2}},  \nonumber \\
\! \! {\rm{B}}:  V_{\rm{fit}}(r)&=&\sum_{i=1,2,3}a_i e^{-(r/b_i)^2}.
\end{eqnarray}

The fit A is motivated by the TPE tail at long distance with an overall strength proportional to $m_{\pi}^4$~\cite{TarrusCastella:2018php},
  while the fit B is  a purely  phenomenological Gaussian form  for comparison.
  In fit A, we consider two types of  form factors commonly used in the $NN$ potentials: (i) the Nijmegen-type form factor $f_{\rm erfc}(r;b_3)$ \cite{Stoks:1994wp}, and (ii) the Argonne-type  form factor $f_{\rm exp}(r;b_3)$~\cite{Wiringa1995}. They are defined as
\begin{eqnarray}
f_{\rm erfc}(r;b_3) &=& \left[  {\rm erfc}\left( \frac{m_{\pi}}{\Lambda}-\frac{\Lambda r}{2}  \right) \right. \nonumber \\
&~~&\left.- e^{2m_{\pi}r} {\rm erfc}\left(  \frac{m_{\pi}}{\Lambda}+\frac{\Lambda r}{2}   \right)  \right]^2/4, \nonumber\\
f_{\rm exp}(r;b_3) &= & \left(1-e^{-(r/b_3)^{2}}\right)^2.
\end{eqnarray}
Here $\Lambda = 2/b_3$ and  ${\rm erfc}(x)= \frac{2}{\sqrt{\pi}} \int_x^{\infty} e^{-z^2} dz$. 
The Nijmegen-type form factor is motivated by the exponential-type regularization of the pion propagator in the momentum space, $1/(k^2+m_{\pi}^2)  \rightarrow e^{-(k/\Lambda)^2 } /(k^2+m_{\pi}^2 )$. 
We refer to fit A with (i) and (ii) as fit A$_{\rm erfc}$ and fit A$_{\rm exp}$, respectively.
The pion mass in fit A$_{\rm erfc, exp}$ is taken to be  $m_\pi=146.4$ MeV, and the fit range is  chosen as  $0 < r < 3.0$ fm.
We found that all fits  provide an equally good result
 ($\chi^2_{\rm d.o.f}=$0.3-0.4) and are stable against the choice of $t$.
 In Table~\ref{tab-prm} we show the fit results for $t/a=14$, which are expected to have least contamination from the inelastic states. 
Changing the fit range of the potential to  $0.1<r<2.5$ fm does not affect the results within statistical errors.
Also we found that the simple fitting functions  such as the 
Yukawa  form $\sim  - \frac{\exp (-\mu r)}{r}$ ~\cite{Brodsky1990,Gao2001} and  the van der Waals (Casimir-Polder) form $\sim - \frac{1}{r^k}$ with $k=6$ ($7$)~\cite{Appelquist1978}  cannot reproduce the lattice data.
 
\begin{table}[hbtp]
\caption{The fit parameters in Eq.~(\ref{eq-fit}) with statistical errors quoted in the parentheses at $t/a=14$.
The fit range is $0<r<3.0$ fm. In $a_3 m_\pi^{4n}$, we take  $n=1$ and $n=0$ for fit A and B, respectively.
A$_{\rm erfc}$ (A$_{\rm exp}$) denotes fit A with the Nijmegen-type (Argonne-type) form factor.}
\begin{tabular}{cccc}
  \hline\hline
    fit~~&A$_{\rm erfc}$~~~&A$_{\rm exp}$~~~&B\\
  \hline
    $a_1$ [MeV] ~~~&-376(20)~~~&-371(27)~~~&-371(19)\\
    $b_1$ [fm] ~~~&0.14(1)~~~&0.13(1)~~~&0.15(3)\\
    $a_2$ [MeV] ~~~&306(122)~~~&-119(39)~~~&-50(35) \\
    $b_2$ [fm] ~~~&0.46(4)~~~&0.30(5)~~~ &0.66(61)\\
    $a_3 m_\pi^{4n}$ [MeV$\cdot{\rm fm}^{2n}$]~~~&-95(13)~~~&-97(14)~~~&-31(53)\\
    $b_3$ [fm] ~~&0.41(7)~~~&0.63(4)~~~&1.09(41) \\
  \hline\hline
\end{tabular} \label{tab-prm}
\end{table}

\begin{figure}[hbtp]
  \centering
  \includegraphics[width=8cm]{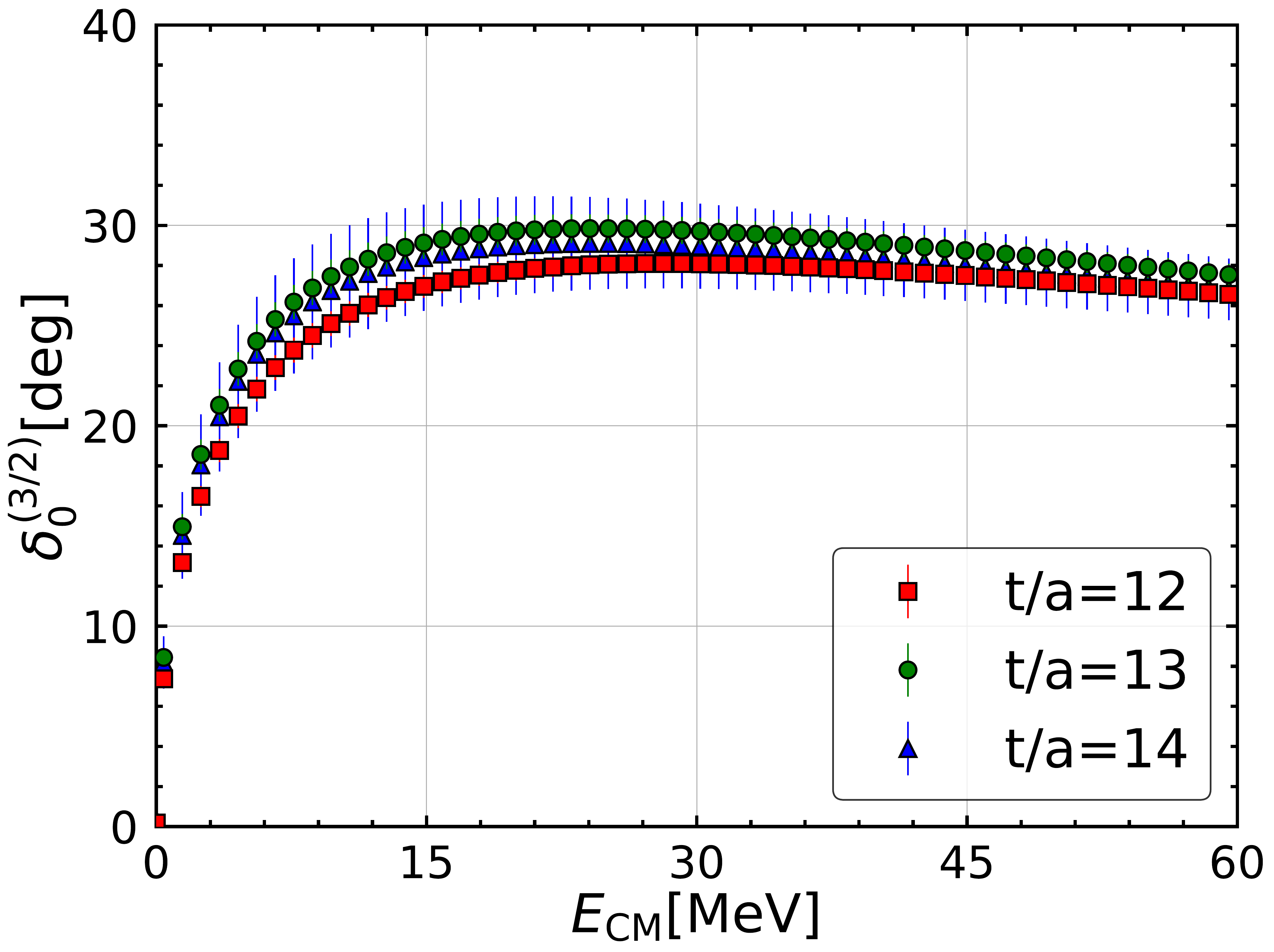}
  \caption{
    The $N$-$\phi$ scattering phase shifts $\delta^{(3/2)}_0$ in the $^4S_{3/2}$ channel obtained from $V_{\rm fit}(r)$ at $t/a=12$  (red squares),  $13$ (green circles), and $14$ (blue triangles).
  } \label{Fig3}
\end{figure}

Figure~\ref{Fig3} shows the $N$-$\phi$ scattering phase shifts $\delta^{(3/2)}_0$ in the $^4S_{3/2}$ channel as a function of the center of mass kinetic energy $E_{\mathrm{CM}}=\sqrt{m_N^2+{k}^2}+\sqrt{m_\phi^2+{k}^2}-(m_N+m_\phi)$ obtained by using $V_{\rm fit}(r)$ with the fit A$_{\rm erfc}$. 
The scattering phase shifts from different $t$ are consistent with each other within statistical errors.
 The scattering length $a^{(3/2)}_0$ and the effective range $r^{(3/2)}_{\rm eff}$ can be extracted from the 
 effective range expansion for small $k$ as
\begin{equation}
k\cot\delta^{(3/2)}_0(k)=-\frac{1}{a^{(3/2)}_0}+\frac12 r^{(3/2)}_{\rm eff}k^2 +O(k^4).
\end{equation} 

 In Table~\ref{tab-scattering}, $a^{(3/2)}_0$ and $r^{(3/2)}_{\rm eff}$ are shown for the present  pion mass $m_{\pi}=146.4$ MeV; 
 the central values and the statistical errors of about 15\% are obtained from the data at $t/a=14$ with the fit A$_{\rm erfc}$,  while 
  the systematic errors of about 25\% in the second parentheses are estimated by comparing results for $t/a=12$-14 with A$_{\rm erfc, exp}$ and B.
Other possible systematic errors are as follows: (i) The finite volume effect, which is expected to be small as  
$\exp (-2m_{\pi} (L/2)a) \lesssim 0.3\%$ due to the large volume; (ii) The finite cutoff effect, which is also expected to be small as 
$O((a\Lambda_{\rm QCD})^2)\sim O(1)\%$ due to the nonperturbative $O(a)$-improvement;
(iii) As an alternative estimate of the cutoff effect, we remove the potential at $r < 0.1$ fm, and  found that the 
scattering parameters  change only $\sim 2\%$; 
(iv) The effect  of $s\bar{s}$ annihilation  is known to  be less than 1\% correction to the 
 $\phi$-meson mass and the mixing to non-$s\bar{s}$ mesons \cite{Dudek2013PRD,Chakraborty:2017hry}.
 Assuming that  the $s\bar{s}$ annihilation effect on $R(\bm{r},t)$ is the similar magnitude of about 1\%, 
the resultant systematic error to the final scattering parameters is found to be less than 1\%.

\begin{table}[tbhp]
\caption{The scattering length $a^{(3/2)}_0$ and the effective range $r^{(3/2)}_{\rm eff}$ obtained by using $V_{\rm fit}(r)$ at $m_\pi=146.4$ MeV
 with statistical and systematic errors. Estimated central values using  a model-dependent extrapolation  of $V_{\rm fit}(r)$ to $m_\pi=138.0$ MeV are 
  also shown for comparison.}
\begin{tabular}{ccc}
  \hline\hline
    $m_\pi$ [MeV]~&$a^{(3/2)}_0$ [fm]~~~&$r^{(3/2)}_{\rm eff}$ [fm]\\
  \hline
    146.4 ~~~&$-1.43(23)_{\rm stat.}\left(^{+36}_{-06}\right)_{\rm syst.}$~~~&$2.36(10)_{\rm stat.}\left(^{+02}_{-48}\right)_{\rm syst.}$\\
      138.0 ~~~&$\simeq -1.25$~~~&$\simeq 2.49$\\

 \hline\hline
\end{tabular} \label{tab-scattering}
\end{table}

To estimate how the scattering parameters change toward the physical quark mass,
we keep $a_{1,2,3}$ and $b_{1,2,3}$ in $V_{\rm fit}(r)$ fixed in fit A$_{\rm erfc, exp}$ and smoothly change the long-range potential by taking the isospin-averaged physical pion mass $m_{\pi} =138.0$ MeV  in the region where the TPE is dominated ($r>1.0$ fm from Fig.\ref{Fig2}).
By calculating the scattering phase shifts with such a potential with the physical masses of  $\phi$ and $N$, we obtain estimated values of  $a^{(3/2)}_0$ and $r^{(3/2)}_{\rm eff}$ for  $m_{\pi} =138.0$ MeV in Table~\ref{tab-scattering}. 
 Although the range of the TPE is increased by the smaller pion mass, the characteristic $m_{\pi}^4$ behavior  of the TPE strength makes the overall attraction weaker.
Note that this is only a model-dependent qualitative estimate and needs to be confirmed by future physical-point simulations.

\begin{figure}[tbhp]
  \centering
  \includegraphics[width=8cm]{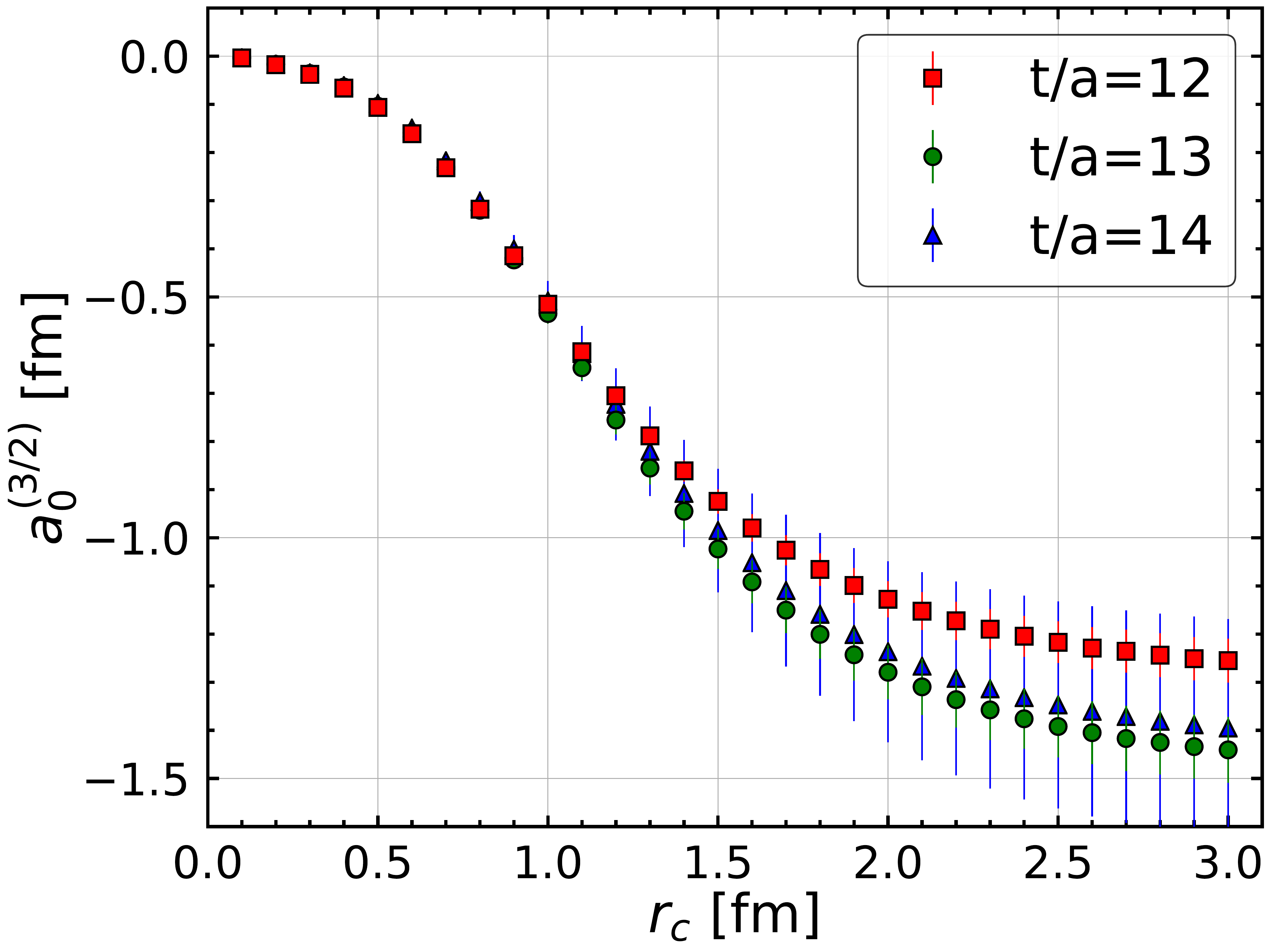}
  \caption{
    The scattering length $a^{(3/2)}_0$ obtained from $V(r;r_{\rm c})=\theta (r_{\rm c}-r)  V_{\rm fit}(r)$ as a function of  cutoff length $r_{\rm c}$ at $t/a=12$  (red squares),  $13$ (green circles), and $14$ (blue triangles).
  } \label{Fig4}
\end{figure}

Since we do not have reliable information on the $N\phi(^2S_{1/2})$ potential from lattice QCD at the moment due to the effect of the open channels,
a comparison of our results with spin-averaged scattering parameters should be made with caution. With this reservation in mind, our $a^{(3/2)}_0$ is found to be one or two orders of magnitude larger than the previous theoretical results in QCD sum rules~\cite{Koike1997,Klingl1997kf}.
Such a discrepancy may be due to the difficulty of obtaining the long-range TPE contribution from the low-order truncation of the operator product expansion in QCD sum rules.  In fact,  the magnitude of $a^{(3/2)}_0$ becomes considerably smaller when the long-range potential is cut off. 
Shown in Fig.~\ref{Fig4} is $a^{(3/2)}_0$ as a function of cutoff length $r_{\rm c}$ obtained by the potential, 
$  V(r;r_{\rm c}) = \theta (r_{\rm c}-r)  V_{\rm fit}(r)$ with the fit A$_{\rm erfc}$.
Considerable decrease of  $|a^{(3/2)}_0|$ from $1.43$ fm at $r_{\rm c}=\infty$ to about $0.1$ fm at $r_{\rm c}=0.5$ fm can be seen.

\section{Summary}\label{sec-V}
In this paper, we present a first lattice QCD calculation on the interaction of the  $N$-$\phi$ system in the $^4S_{3/2}$ 
channel based on the  ($2+1$)-flavor simulations with nearly physical quark masses.
The interaction potential in the  $N\phi(^4S_{3/2})$ channel is extracted from lattice data of the 
hadronic spacetime correlation using the HAL QCD method.
The potential is found to be attractive for all distances and appears to be a combination of an attractive core at short distances and a two-pion exchange (TPE) tail at long distances ($r >$ 1 fm).
The latter is well fitted by the characteristic form of the TPE obtained by the interaction of a color-dipole and the nucleon.
The scattering parameters obtained from our potential at $m_{\pi}=146.4$ MeV is summarized in Table \ref{tab-scattering}.
By examining the potential fitted to the lattice data, we find that the scattering length $a^{(3/2)}_0$ is sensitive to the length scale of $r > 0.5$ fm.  
Also, we suggest that the $N$-$\phi$ attraction could be weaker at the physical pion mass due to the characteristic $m^4_{\pi}$ dependence of the strength of the TPE.
 
Our $a^{(3/2)}_0$ is substantially larger in magnitude than the previous calculations of the spin-averaged $a_0$ using QCD sum rules but is comparable to the spin-averaged $a_0$  by ALICE Collaboration within the error bar ~\cite{ALICE2021}.
Also, our $r^{(3/2)}_{\rm eff}$ is about three times smaller than the spin-averaged $r_{\rm eff}$ by ALICE Collaboration.
To make a solid comparison between theory and experiments, we need to extract  
  complex-valued scattering parameters  in the $^2S_{1/2}$ and $^4S_{3/2}$  channels
through the coupled-channel analysis of the data from physical-point simulations.
The present lattice QCD study near the physical point provides a first step to exploring the interaction of $s\bar{s}$ with the nucleon
 from the first principles.  The heavier system such as $c\bar{c}$ interacting with the nucleon pioneered in ~\cite{Kawanai:2010ev,Sugiura:2019pye}
  is also an important problem to be studied. 

\begin{acknowledgments}
We thank members of the HAL QCD Collaboration for stimulating discussions. 
Y. L. thanks Xu Feng for the helpful discussions.  We thank members of the PACS Collaboration for the gauge configuration generation conducted on the K computer at RIKEN.
The lattice QCD measurements have been performed on Fugaku and HOKUSAI supercomputers at RIKEN.
We thank ILDG/JLDG \cite{ldg}, which serves as essential infrastructure in this study.
This work was partially supported by HPCI System Research Project (hp120281, hp130023, hp140209, hp150223, hp150262, hp160211, hp170230, hp170170, hp180117, hp190103, hp200130, hp210165, and hp210212),  the National Key R\&D Program of China (Contracts No. 2017YFE0116700 and No. 2018YFA0404400), the National Natural Science Foundation of China (Grants No. 11935003, No. 11975031, No. 11875075, and No. 12070131001), the JSPS (Grants No. JP18H05236, No. JP16H03978, No. JP19K03879, No. JP18H05407, and No. JP21K03555), the MOST-RIKEN Joint Project ``{\it Ab initio} investigation in nuclear physics'', ``Priority Issue on Post-K computer'' (Elucidation of the Fundamental Laws and Evolution of the Universe), ``Program for Promoting Researches on the Supercomputer Fugaku'' (Simulation for basic science: from fundamental laws of particles to creation of nuclei), and Joint Institute for Computational Fundamental Science (JICFuS).

\end{acknowledgments}

\appendix

\section{FINITE-VOLUME SPECTRAL ANALYSIS}\label{sec-VI}

\begin{figure}[htbp]
  \centering
  \includegraphics[width=8cm]{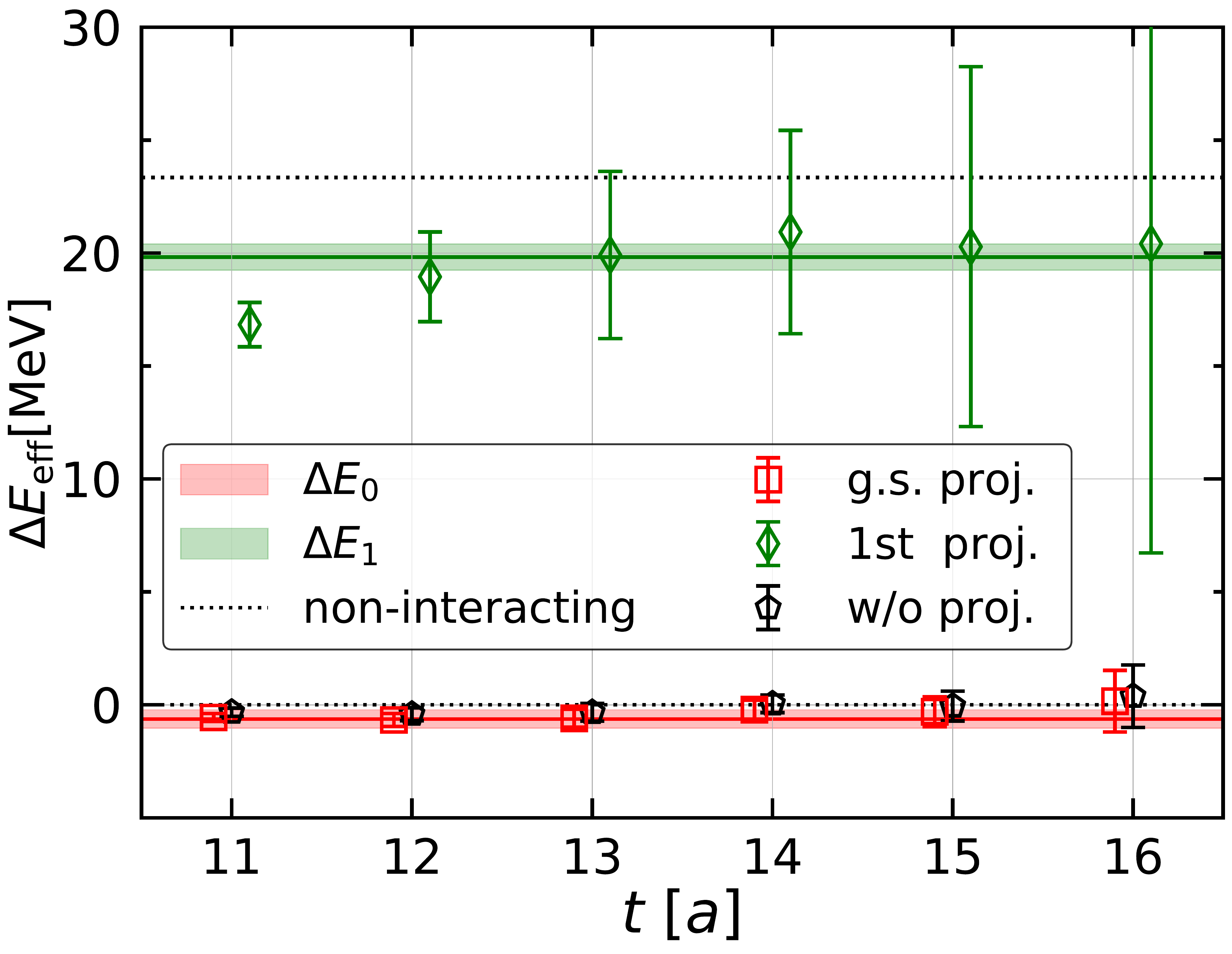}
  \caption{
 The effective energies $\Delta E_{0,1}$ from the LO potential $V(r)$ (colored bands) and $\Delta E^{\rm eff}_{0,1}$ from the projected temporal correlation functions (colored points). Black pentagons represent the effective energy extracted from the temporal correlation function without projection, i.e. $R(t)=\sum_{\bm r}R(\bm r, t)$. The black dotted lines are $\Delta E_{0,1}$ for a noninteracting system.
  } \label{Supp-Fig2}
\end{figure}

Let us consider the truncation error of the derivate expansion. For this purpose,
 we construct a Hamiltonian $H$ in a three-dimensional lattice box with the LO potential $V(r)$ ~\cite{Iritani2019Jhep,Lyu2022},
\begin{equation}
 H= -\frac{\nabla^2}{2\mu} + V(r), \quad H\psi_n(\bm r)=\varepsilon_n \psi_n(\bm r).
\end{equation}
Here $\varepsilon_n$ is related to $\Delta E_n=\sqrt{m_N^2+2\mu\varepsilon_n}+\sqrt{m_\phi^2+2\mu\varepsilon_n}-(m_N+m_\phi)$.
From the NBS wave function $\psi_n(\bm r)$, one can construct an optimized sink operator as a projection to each $n$th state,
\begin{equation}
  S_n=\sum_{\bm r}\psi^\dagger_n(\bm r) \left[\sum_{\bm x}N(\bm r+\bm x,t)\phi(\bm x,t)\right],
\end{equation}
which is expected to overlap largely to the $n$th state.
The temporal correlation function with such an optimized sink operator can be obtained as
\begin{equation}
 R_n(t)=\sum_{\bm r}\psi^\dagger_n (\bm r)R(\bm r,t)= a_ne^{-{(\Delta E_n)t}} + O(e^{-(\Delta E^*t)}).
\end{equation}
The effective energy for the $n$th state can be defined from $R_n(t)$ as
\begin{equation}
 \Delta E^{\rm eff}_n= \frac{1}{a}\ln\left[\frac{R_n(t)}{R_n(t+1)}\right].
\end{equation}
Thus, by comparing $\Delta E_n$ from the LO potential $V(r)$ and $\Delta E^{\rm eff}_n$ from the projected temporal correlation function, we can make a highly nontrivial check on the systematic errors in the LO potential $V(r)$.

Shown in Fig.~\ref{Supp-Fig2} are $\Delta E_{0,1}$ from the LO potential $V(r)$ (colored bands), and $\Delta E^{\rm eff}_{0,1}$ from the projected temporal correlation functions (colored points).
We also show the effective energy from the temporal correlation function without projection, i.e. $R(t)=\sum_{\bm r}R(\bm r, t)$, for comparison (black points).
We found that $\Delta E_{0,1}$ and $\Delta E^{\rm eff}_{0,1}$ are consistent with each other within statistical errors, which indicates that the systematic errors in the LO potential $V(r)$ are well under control.

\section{TIME DEPENDENCE OF THE $N$-$\phi$ POTENTIAL}\label{sec-VII}

\begin{figure}[htbp]
  \centering
  \includegraphics[width=8cm]{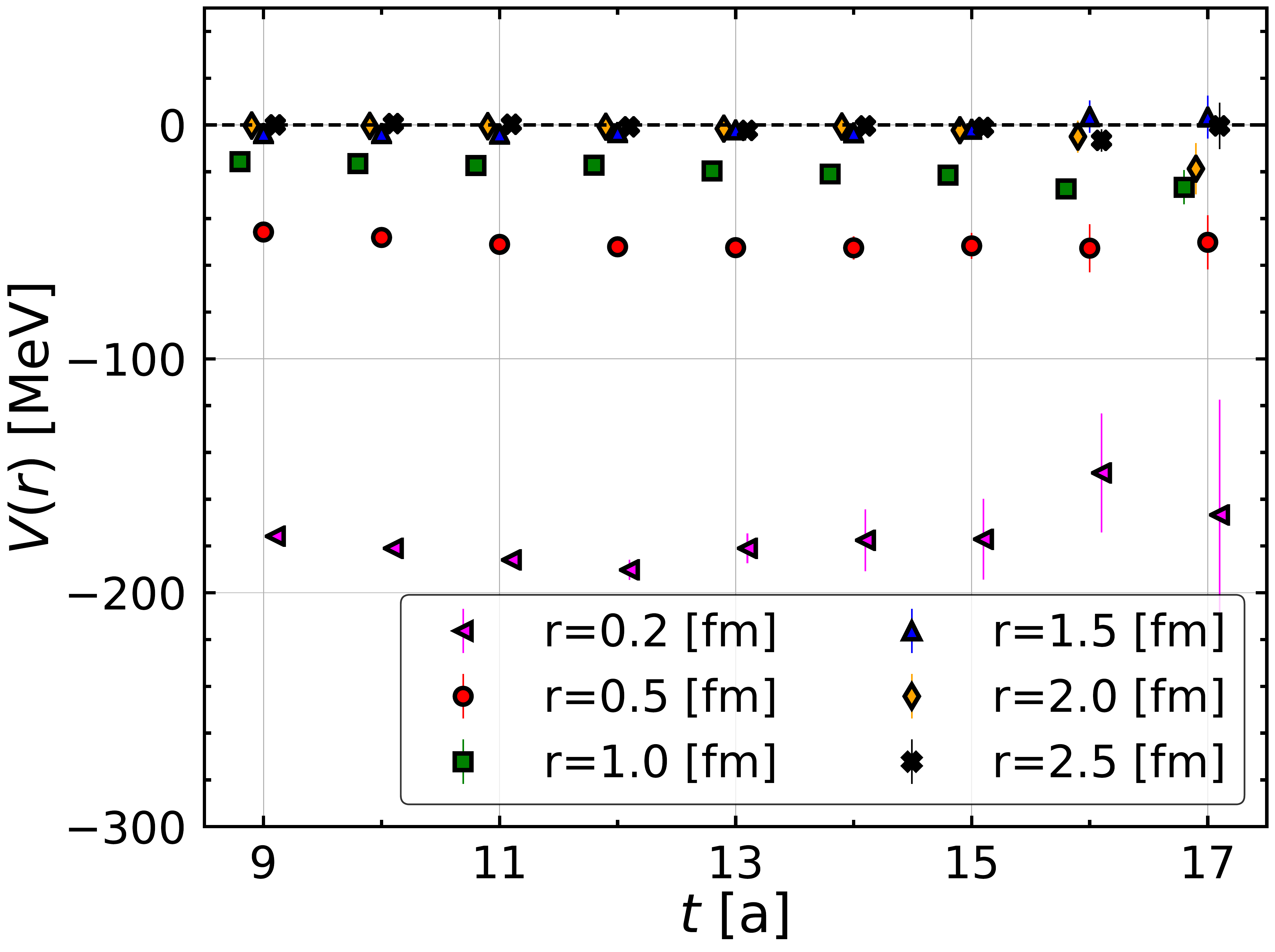}
  \caption{
 The $t$ dependence of the $N$-$\phi$ potential $V(r)$ in the ${^4S_{3/2}}$ channel for several distances $r$. For visibility, data are a little shifted horizontally. }
\label{Supp-Fig1}
\end{figure}

In Fig.~\ref{Supp-Fig1}, we show the $t$ dependence of  $V(r)$ for several distances $r=0.2$, $0.5$, $1.0$, $1.5$, $2.0$, and $2.5$ fm in a wide range of $t$, $9\leq t/a\leq 17$.
We found the potential at given $r$ varies slowly with $t$: 
This provides alternative evidence that the truncation error due to the LO approximation is small.  Also it indicates that
the elastic scattering states in $N\phi(^4S_{3/2})$ play a dominant role at these Euclidean times. 
In other words, if $N\phi(^4S_{3/2})$ strongly couples to open channels such as $\Lambda K(^2D_{3/2})$, $\Sigma K(^2D_{3/2})$, $\Lambda\pi K$, and $\Sigma\pi K$, the potential at given $r$ would decrease monotonically with $t$ increasing.



\

\end{document}